\documentclass{elsart}
\usepackage[latin1]{inputenc}
\usepackage{amssymb}
\date{}

\newtheorem{The}{Theorem}[section]

\newtheorem{Pro}[The]{Proposition}
\newtheorem{Deff}[The]{Definition}
\newtheorem{Lem}[The]{Lemma}
\newtheorem{Rem}[The]{Remark}

\newcommand{\fa}{\forall}
\newcommand{\Ga}{\Gamma}
\newcommand{\Gas}{\Gamma^\star}
\newcommand{\Si}{\Sigma}

\newcommand{\Sis}{\Sigma^\star}
\newcommand{\ra}{\rightarrow}
\newcommand{\hs}{\hspace{12mm}

\noi}

\newcommand{\de}{deterministic }
\newcommand{\tla}{\twoheadleftarrow}
\newcommand{\lra}{\leftrightarrow}
\newcommand{\la}{language}
\newcommand{\ite}{\item}

\newcommand{\ol}{ $\omega$-language}
\newcommand{\orl}{ $\omega$-regular language}
\newcommand{\om}{\omega}
\newcommand{\nl}{\newline}
\newcommand{\noi}{\noindent}
\newcommand{\proo}{\noi {\bf Proof.} }
\newcommand {\ep}{\hfill $\square$}

\begin{document}
\begin{frontmatter}
\title{\bf ON OMEGA CONTEXT FREE LANGUAGES WHICH ARE BOREL SETS OF INFINITE RANK}

\author{Olivier Finkel\corauthref{cor}}
\corauth[cor]{Corresponding author}
\ead{finkel@logique.jussieu.fr }

\address{ Equipe de Logique Math\'ematique \\CNRS  et 
Universit\'e Paris 7,  U.F.R. de Math\'ematiques \\  2 Place Jussieu 75251 Paris
 cedex 05, France.}
 
\begin{abstract}
\noi This paper is a continuation of the study of topological properties of omega
 context free languages ($\om$-CFL). We proved in 
 [Topological Properties of Omega Context Free Languages, 
Theoretical Computer Science, Volume 262 (1-2), 2001, p. 669-697] 
 that the class of $\om$-CFL
 exhausts the finite ranks of the Borel hierarchy, and in 
[Borel Hierarchy and Omega Context Free Languages, 
 Theoretical Computer Science, to appear]  that there 
exist some $\om$-CFL which are analytic but non Borel sets. 
We prove here that there exist some  omega
 context free languages which are Borel sets of infinite (but not finite) rank, 
giving  additional answer to questions of Lescow and Thomas 
 [Logical specifications of infinite computations,
 In:"A Decade of Concurrency" (J. W. de Bakker et al., eds), Springer LNCS 803 (1994), 583-621].
\end{abstract}

\end{frontmatter}

\tableofcontents

\section{Introduction}

Since J.R. B\"uchi studied the \ol s recognized   by finite   automata to 
prove 
the decidability of the monadic second order theory of one successor
 over the integers \cite{bu60a}
 the so called $\om$-regular languages have been intensively
 studied.
See \cite{tho} and \cite{pp} for many results and references. 

 Pushdown automata are a natural extension of finite automata.  R. S.
 Cohen and A. Y. Gold \cite{cg} , \cite{cgdet} 
  and M. Linna \cite{lin76}  studied the \ol s accepted by omega 
pushdown automata,
 considering various acceptance conditions for omega words.
It turned out that the omega languages accepted by  omega pushdown 
automata were also those
 generated by context free grammars where infinite derivations are
 considered, also studied by 
M. Nivat \cite{ni77} \cite{ni78}  and L. Boasson and M. Nivat \cite{bn}.  These 
languages were then 
called the 
 omega context free  languages ($\om$-CFL).
See also Staiger's paper \cite{sta} for a survey of general theory of
\ol s, including 
more powerful accepting devices, like Turing machines, and the fundamental study 
of J. Engelfriet and H. J. Hoogeboom on {\bf X}-automata, i.e. finite automata equipped with 
a storage type {\bf X}, reading infinite words \cite{eh}.

  Topology is a useful tool for classifying \ol s by the study of  their complexity, 
particularly with regard to the Borel hierarchy.

  McNaughton's Theorem implies that \orl s (\ol s accepted by \de Muller automata)
 are boolean 
combination of ${\bf \Pi^0_2}$-sets, \cite{rmc}.  
Topological properties of \orl s were first
 studied by L. H. Landweber in \cite{la} where he characterized  
 \orl s  in a given Borel class.

 J. Engelfriet and H. J. Hoogeboom proved that all \ol s  accepted by {\it  \de }
 {\bf X}-automata
with a Muller acceptance condition are also 
boolean combinations of ${\bf \Pi^0_2}$-sets hence 
(${\bf \Si^0_3}\cap {\bf \Pi^0_3}$)-sets. 

 When considering {\it  non \de} finite machines, as {\bf X}-automata, a natural 
question,  posed by H. Lescow and W. Thomas in \cite{lt}, 
now arises:  what is the topological complexity of \ol s accepted 
by automata equipped with a given storage type {\bf X}? Are they all Borel sets of finite rank, 
Borel sets, analytic sets? 

 It is well known that every \ol~accepted by a Turing machine (hence also by a 
{\bf X}-automaton) with a Muller acceptance condition is an analytic set \cite{sta} 
(i.e. is obtained as a continuous image of a Borel set or as the projection 
of a Borel set \cite{mos}). 

 We consider in this paper the storage type "{\it pushdown}". We  
pursue  the investigation of topological properties 
of omega context free  languages. We proved that 
the class of $\om$-CFL exhausts the finite ranks of the Borel  hierarchy, giving 
examples of ${\bf \Pi^0_n}$-complete (respectively ${\bf \Si^0_n}$-complete) $\om$-CFL  
for each integer $n\geq 1$,  \cite{fin}. We showed in \cite{finb} that there exist some 
omega context free  languages which are analytic but non Borel sets. There exist such \ol s 
in the form $L^\om$, with $L$ a context free finitary language; this gave an answer to 
questions of D. Niwinski and P. Simonnet about omega powers of finitary 
languages \cite{niw} \cite{sim}. 

But the  question was still open whether there exist some  omega
 context free languages which are Borel sets of infinite  rank.  

 We answer to this question in this paper giving examples of $\om$-CFL which 
are Borel sets of infinite rank.

  The paper is organized as follows. 
In sections 2 and 3, we first review some above definitions and results about $\om$-regular,
 $\om$-context free languages, and topology.
\nl 
In section 4  we introduce   the operation of  exponentiation of sets defined by 
J.  Duparc in his recent study of the Wadge hierarchy of 
Borel sets, which is a great refinement of the Borel hierarchy  \cite{dup}, and 
recall preceding results of \cite{fin}.   
\nl 
In section 5, we prove our main result about  $\om$-CFL, using an  
iteration of Duparc's operation and  
give additional  answer to questions of  W. Thomas and H. Lescow \cite{lt}.

\section{$\om$-regular and  $\om$-context free  \la s}

We assume the reader to be familiar with the theory of formal \la s and 
of \orl s, see for example \cite{hu69} ,\cite{tho}.
We first recall some of the definitions and results concerning $\om$-regular 
and $\om$-context free  \la s and omega pushdown automata as presented in \cite{tho} 
\cite{cg} 
\cite{cgdet}.
\nl
When $\Si$ is a finite alphabet, a finite string (word) over $\Si$ is any 
sequence $x=x_1\ldots x_k$ , where $x_i\in\Sigma$ 
for $i=1,\ldots ,k$ ,and  $k$ is an integer $\geq 1$. The length
 of $x$ is $k$, denoted by $|x|$ .
\nl
 If  $|x|=0$ , $x$ is the empty word denoted by $\lambda$. 
\nl we write $x(i)=x_i$  and $x[i]=x(1)\ldots x(i)$ for $i\leq k$ and $x[0]=\lambda$.
\nl $\Sis$  is the set of finite words over $\Sigma$.
\nl  The first infinite ordinal is $\om$.
\nl An $\om$-word over $\Si$ is an $\om$ -sequence $a_1\ldots a_n\ldots  $, where 
$a_i \in\Sigma , \fa i\geq 1$.
\nl When $\sigma$ is an $\om$-word over $\Si$, we write
 $\sigma =\sigma(1)\sigma(2)\ldots \sigma(n) \ldots $
\nl and $\sigma[n]=\sigma(1)\sigma(2)\ldots \sigma(n)$ 
the finite word of length n, prefix of $\sigma$.
\nl The set of $\om$-words over  the alphabet $\Si$ is denoted by $\Si^\om$.
\nl An  $\om$-language over an alphabet $\Sigma$ is a subset of  $\Si^\om$.

 The usual concatenation product of two finite words $u$ and $v$ is 
denoted $u.v$ (and sometimes just $uv$). This product is extended to the product of a 
finite word $u$ and an $\om$-word $v$: the infinite word $u.v$ is then the $\om$-word such that:
\nl $(u.v)(k)=u(k)$  if $k\leq |u|$ , and 
\nl $(u.v)(k)=v(k-|u|)$  if $k>|u|$.

 For $V\subseteq \Sis$, $V^\om = \{ \sigma =u_1\ldots u_n \ldots \in \Si^\om /  u_i\in V, \fa i\geq 1 \}$
is the $\om$-power of $V$.
\nl For $V\subseteq \Sis$, the complement of $V$ (in $\Sis$) is $\Sis - V$ denoted $V^-$.
\nl  For a subset $A\subseteq \Si^\om$, the complement of $A$ is 
$\Si^\om - A$ denoted $A^-$.

 The prefix relation is denoted $\sqsubseteq$: the finite word $u$ is a prefix of the finite 
word $v$
(denoted $u\sqsubseteq v$) if and only if there exists a (finite) word $w$ such that $v=u.w$.
\nl This definition is extended to finite words which are prefixes of $\om$-words:
\nl the finite word $u$ is a prefix of the $\om$-word $v$ (denoted $u\sqsubseteq v$) 
iff there exists an $\om$-word $w$ such that $v=u.w$.

\begin{Deff} : A finite state machine (FSM) is a quadruple $M=(K,\Si,\delta, q_0)$, where $K$ 
is a finite set of states, $\Sigma$ is a finite input alphabet, $q_0 \in K$ is the initial state
and $\delta$ is a mapping from $K \times   \Si$ into $2^K$ . A FSM is called deterministic
 (DFSM) iff :
$\delta : K \times  \Si \ra K$.
\nl 
A B\"uchi automaton (BA) is a 5-tuple $M=(K,\Si,\delta, q_0, F)$ where
  $M'=(K,\Si,\delta, q_0)$
is a finite state machine and $F\subseteq K$ is the set of final states.
\nl
A Muller automaton (MA) is a 5-tuple $M=(K,\Si,\delta, q_0, \mathcal{F})$ where
$M'=(K,\Si,\delta, q_0)$ is a FSM and $\mathcal{F}\subseteq 2^K$ is the collection of 
designated state sets.
\nl
A B\"uchi or Muller automaton is said \de if the associated FSM is deterministic.
\nl
Let $\sigma =a_1a_2\ldots a_n\ldots$ be an  $\om$-word over $\Si$.
 \nl 
A sequence of states $r=q_1q_2\ldots q_n\ldots$  is called an (infinite) run of $M=(K,\Si,\delta, q_0)$ on $\sigma$, 
starting in state $p$, iff:
1) $q_1=p$  and 2) for each $i\geq 1$, $q_{i+1} \in \delta( q_i,a_i)$.
\nl
In case a run $r$ of $M$ on $\sigma$ starts in state $q_0$, we call it simply "a run of $M$ 
on $\sigma$ " .
\nl
For every (infinite) run $r=q_1q_2\ldots q_n\ldots $ of $M$, $In(r)$ is the set of
states in $K$ entered by $M$ infinitely many times during run $r$:
\nl
$In(r)= \{ q\in K  / \{i\geq 1  / q_i=q\} $   is infinite $\}   $.
\nl For $M=(K,\Si,\delta, q_0, F)$ a BA ,
the \ol~ accepted by $M$ is
$L(M)= \{  \sigma\in\Si^\om$ / there exists a  run r
 of M on $\sigma$ such that $In(r) \cap F \neq\emptyset \}$.
\nl For $M=(K,\Si,\delta, q_0, F)$ a MA, the  \ol~ accepted by $M$ is 
$L(M)= \{  \sigma\in\Si^\om$ / there exists a  run r
 of M on $\sigma$ such that $In(r) \in \mathcal{F} \}$.
\end{Deff}

\noi The classical result of R. Mc Naughton \cite{rmc}  established that the expressive
 power of \de MA (DMA) is equal to the expressive power of non \de MA
(NDMA) which is also equal to the expressive power of non \de BA (NDBA) .
\nl There is  also a characterization of the \la s accepted by MA by means 
of the "$\om$-Kleene closure" which we give now the definition:

\begin{Deff}
For any family L of  finitary \la s over the alphabet $\Si$, the $\om$-Kleene closure
of L, is : $$\om-KC(L) = \{ \cup_{i=1}^n U_i.V_i^\om  /  U_i, V_i \in L , \fa i\in [1, n] \}$$
\end{Deff}

\begin{The}
  For any \ol~ $L$, the following conditions are equivalent:
\begin{enumerate}
\ite   $L$ belongs to $\om-KC(REG)$ , where $REG$ is the class of (finitary)
 regular languages.
\ite   There exists a DMA  that accepts $L$.
\ite   There exists a MA  that accepts $L$.
\ite     There exists a BA  that accepts $L$.
\end{enumerate}

\noi An \ol~ $L$ satisfying one of the conditions of the above Theorem is called 
 an \orl . 
\nl The class of \orl s will
 be denoted by $REG_\om$.
\end{The}

\noi We now define the pushdown machines and the classes of  $\om$-context free  \la s.

\begin{Deff}
A pushdown machine (PDM) is a 6-tuple $M=(K,\Si,\Ga, \delta, q_0, Z_0)$, where $K$ 
is a finite set of states, $\Sigma$ is a finite input alphabet, $\Gamma$ is a 
finite pushdown alphabet,
 $q_0\in K$ is the initial state, $Z_0 \in\Ga$ is the start symbol, 
and $\delta$ is a mapping from $K \times (\Si\cup\{\lambda\} )\times \Ga $ to finite subsets of
$K\times \Gas$ . 
\nl
If  $\gamma\in\Ga^{+}$ describes the pushdown store content, 
the leftmost symbol will be assumed to be on " top" of the store.
A configuration of a PDM is a pair $(q, \gamma)$ where $q\in K$ and  $\gamma\in\Gas$.\nl
For $a\in \Si\cup\{\lambda\}$, $\gamma,\beta\in\Ga^{\star}$
and $Z\in\Ga$, if $(p,\beta)$ is in $\delta(q,a,Z)$, then we write
$a: (q,Z\gamma)\mapsto_M (p,\beta\gamma)$.\nl
$\mapsto_M^\star$ is the transitive and reflexive closure of $\mapsto_M$.
(The subscript $M$ will be omitted whenever the meaning remains clear).
\nl
Let $\sigma =a_1a_2\ldots a_n\ldots$ be an  $\om$-word over $\Si$. 
An infinite sequence of configurations $r=(q_i,\gamma_i)_{i\geq1}$ is called 
a  run of $M$ on $\sigma$, starting in configuration $(p,\gamma)$, iff:
\begin{enumerate}
\ite $(q_1,\gamma_1)=(p,\gamma)$

\ite  for each $i\geq 1$, there exists $b_i\in\Si\cup\{\lambda\}$ 
satisfying $b_i: (q_i,\gamma_i)\mapsto_M(q_{i+1},\gamma_{i+1} )$
such that either ~  $a_1a_2\ldots a_n\ldots =b_1b_2\ldots b_n\ldots$ 
\nl or ~  $b_1b_2\ldots b_n\ldots$ is a finite prefix of ~ $a_1a_2\ldots a_n\ldots$
\end{enumerate}
\noi The run $r$ is said to be complete when $a_1a_2\ldots a_n\ldots =b_1b_2\ldots b_n\ldots$ 

 As for FSM, for every such run, $In(r)$ is the set of all states entered infinitely
 often during run $r$.
\nl
A complete run $r$ of $M$ on $\sigma$ , starting in configuration $(q_0,Z_0)$,
 will be simply called " a run of $M$ on $\sigma$ ".
\end{Deff}

\begin{Deff} A B\"uchi pushdown automaton (BPDA) is a 7-tuple
 $M=(K,\Si,\Ga, \delta, q_0, Z_0, F)$ where $ M'=(K,\Si,\Ga, \delta, q_0, Z_0)$
is a PDM and $F\subseteq K$ is the set of final states.
\nl
The \ol~ accepted by $M$ is 
$L(M)= \{  \sigma\in\Si^\om$ / there exists a complete run r
 of M on $\sigma$ such that $In(r) \cap F \neq\emptyset \}$.
\end{Deff}

\begin{Deff} A Muller pushdown automaton (MPDA) is a 7-tuple
 $M=(K,\Si,\Gamma, \delta, q_0, Z_0, \mathcal{F})$ 
where $ M'=(K,\Si,\Gamma, \delta, q_0, Z_0)$
is a PDM and $\mathcal{F}\subseteq 2^K$ is the collection of designated state sets.
\nl
The \ol~ accepted by $M$ is 
$L(M)= \{  \sigma\in\Si^\om$ / there exists a complete run r
 of M on $\sigma$ such that $In(r) \in \mathcal{F} \}$.
\end{Deff}

\begin{Rem} We consider here two acceptance conditions for $\om$-words , 
the B\"uchi  and the Muller acceptance conditions, respectively denoted 2-acceptance 
and 3-acceptance in \cite{la} and in \cite{cgdet} and $(inf, \sqcap)$ and $(inf, =)$ 
in \cite{sta}.
\end{Rem}

\noi R.S. Cohen and A.Y. Gold,  and independently M. Linna, established a characterization 
Theorem for $\om$-CFL:

\begin{The}\label{theokccf}
Let $CFL$ be the class of context free (finitary) languages. Then for any  \ol~ $L$ the following
three conditions are equivalent:
\begin{enumerate}
\ite $L\in \om -KC(CFL)$.
\ite There exists a $BPDA$ that accepts $L$.
\ite There exists a $MPDA$ that accepts $L$.
\end{enumerate}
\end{The}

\noi
In \cite{cg}  are also studied the \ol s generated by $\om$-context free grammars  
and it is shown that each of the conditions 1), 2), and 3) of the above Theorem is 
also equivalent to: 4) $L$ is generated by a context free grammar $G$ by leftmost derivations.
These grammars are also studied in \cite{ni77} , \cite{ni78}.
\nl
Then we can let the following definition:

\begin{Deff}
An \ol~ is an $\om$-context free language ($\om$-CFL) iff it satisfies 
one of the conditions of the above Theorem.
\end{Deff}

\section{Topology}

\noi We assume the reader to be familiar with basic notions of topology which
may be found in \cite{mos} \cite{lt} \cite{sta} \cite{pp} and  with the elementary theory 
of  (countable) ordinals.

Topology is an important tool for the study of \ol s, and leads 
to characterization of several classes of \ol s.
\nl For a finite alphabet $X$, we consider $X^\om$ 
as a topological space with the Cantor topology.
 The open sets of $X^\om$ are the sets in the form $W.X^\om$, where $W\subseteq X^\star$.
A set $L\subseteq X^\om$ is a closed set iff its complement $X^\om - L$ is an open set.
The class of open sets of $X^\om$ will be denoted by ${\bf G}$ or by ${\bf \Si^0_1 }$. 
The class of closed sets will be denoted by ${\bf F}$ or by ${\bf \Pi^0_1 }$. 
Closed sets are characterized by the following:

\begin{Pro}
A set $L\subseteq X^\om$ is a closed set of $X^\om$ iff for every $\sigma\in X^\om$, 

$[\fa n\geq 1,  \exists u\in X^\om$  such that $\sigma (1)\ldots \sigma (n).u \in L]$
 implies that $\sigma\in L$.
\end{Pro}

\noi Define now the next classes of the Borel Hierarchy:

\begin{Deff}
The classes ${\bf \Si_n^0}$ and ${\bf \Pi_n^0 }$ of the Borel Hierarchy
 on the topological space $X^\om$  are defined as follows:
\nl ${\bf \Si^0_1 }$ is the class of open sets of $X^\om$.
\nl ${\bf \Pi^0_1 }$ is the class of closed sets of $X^\om$.
\nl ${\bf \Pi^0_2 }$  or ${\bf G_\delta }$ is the class of countable intersections of 
 open sets of $X^\om$.
\nl  ${\bf \Si^0_2 }$  or ${\bf F_\sigma }$ is the class of countable unions  of 
closed sets of $X^\om$.
\nl And for any integer $n\geq 1$:
\nl ${\bf \Si^0_{n+1} }$   is the class of countable unions 
of ${\bf \Pi^0_n }$-subsets of  $X^\om$.
\nl ${\bf \Pi^0_{n+1} }$ is the class of countable intersections of 
${\bf \Si^0_n}$-subsets of $X^\om$.
\nl The Borel Hierarchy is also defined for transfinite levels.
The classes ${\bf \Si^0_\alpha }$
 and ${\bf \Pi^0_\alpha }$, for a countable ordinal $\alpha$, are defined in the
 following way:
\nl ${\bf \Si^0_\alpha }$ is the class of countable unions of subsets of $X^\om$ in 
$\cup_{\gamma <\alpha}{\bf \Pi^0_\gamma }$.
 \nl ${\bf \Pi^0_\alpha }$ is the class of countable intersections of subsets of $X^\om$ in 
$\cup_{\gamma <\alpha}{\bf \Si^0_\gamma }$.
\end{Deff}

\noi Recall some basic results about these classes, \cite{mos}:

\begin{Pro}
\noi  
\begin{enumerate}
\ite[(a)] ${\bf \Si^0_\alpha }\cup {\bf \Pi^0_\alpha } \subsetneq  
{\bf \Si^0_{\alpha +1}}\cap {\bf \Pi^0_{\alpha +1} }$, for each countable 
ordinal  $\alpha \geq 1$. 
\ite[(b)] $\cup_{\gamma <\alpha}{\bf \Si^0_\gamma }= \cup_{\gamma <\alpha}{\bf \Pi^0_\gamma }
\subsetneq {\bf \Si^0_\alpha }\cap {\bf \Pi^0_\alpha }$, for each countable limit ordinal 
$\alpha$. 
\ite[(c)] A set $W\subseteq X^\om$ is in the class ${\bf \Si^0_\alpha }$ iff its 
complement is in the class ${\bf \Pi^0_\alpha }$. 
\ite[(d)] ${\bf \Si^0_\alpha } - {\bf \Pi^0_\alpha } \neq \emptyset $ and 
${\bf \Pi^0_\alpha } - {\bf \Si^0_\alpha } \neq \emptyset $ hold 
 for every countable  ordinal $\alpha\geq 1$. 
\ite[(e)] For every ordinal $\alpha\geq 1$, the class ${\bf \Si^0_\alpha }$ is closed under 
countable unions and the class ${\bf \Pi^0_\alpha }$ is closed under 
countable intersections.  
\end{enumerate}
\end{Pro}

\noi  We shall say that a subset of $X^\om$ is a Borel set of rank $\alpha$, for 
a countable ordinal $\alpha$,  iff 
it is in ${\bf \Si^0_{\alpha}}\cup {\bf \Pi^0_{\alpha}}$ but not in 
$\bigcup_{\gamma <\alpha}({\bf \Si^0_\gamma }\cup {\bf \Pi^0_\gamma})$.

 There is a nice characterization of ${\bf \Pi^0_2 }$-subsets of $X^\om$. 
First define the notion of $W^\delta$:

\begin{Deff}
For $W\subseteq X^\star$,
 let: 
\nl $W^\delta=\{\sigma\in X^\om / \exists^\om i$ such that $\sigma[i]\in W\}$.
\nl ($\sigma \in W^\delta$ iff $\sigma$ has infinitely many prefixes in $W$).
\end{Deff}

\noi Then we can state the following Proposition:

\begin{Pro}[see \cite{sta}]\label{pi}
A subset $L$ of $X^\om$ is a ${\bf \Pi^0_2 }$-subset of $X^\om$ iff there exists 
a set $W\subseteq X^\star$ such that $L=W^\delta$.
\end{Pro}

\noi For  $X$   a finite set, (and this is also true if $X$ is an infinite 
alphabet)  there are some subsets of $X^\om$ which 
are not Borel sets.
Indeed there exists another hierarchy beyond the Borel hierarchy, which is called the 
projective hierarchy and which is obtained from  the Borel hierarchy by 
successive applications of operations of projection and complementation.
More precisely, a subset $A$ of  $X^\om$ is in the class ${\bf \Si^1_1}$ of {\bf analytic} sets
iff there exists another finite set $Y$ and a Borel subset $B$  of  $(X\times Y)^\om$ 
such that $ x \in A \lra \exists y \in Y^\om $ such that $(x, y) \in B$.
\nl We denote $(x, y)$  the infinite word over the alphabet $X\times Y$ such that
$(x, y)(i)=(x(i),y(i))$ for each  integer $i\geq 0$.
\nl Now a subset of $X^\om$ is in the class ${\bf \Pi^1_1}$ of {\bf coanalytic} sets
iff its complement in $X^\om$ is an analytic set.
\nl The next classes are defined in the same manner, ${\bf \Si^1_{n+1}}$-sets of  
$X^\om$ are projections of ${\bf \Pi^1_n}$-sets and  ${\bf \Pi^1_{n+1}}$-sets 
are the complements of ${\bf \Si^1_{n+1}}$-sets.

 Recall also the notion of completeness with regard to reduction by continuous functions. 
\nl Let $\alpha$ be a countable ordinal. A set $F\subseteq X^\om$ is a ${\bf \Si^0_\alpha}$
 (respectively ${\bf \Pi^0_\alpha}$)-complete set iff for any set $E\subseteq Y^\om$
(with $Y$ a finite alphabet): 
\nl $E\in {\bf \Si^0_\alpha}$ (respectively $E\in {\bf \Pi^0_\alpha}$) 
iff there exists a continuous 
function $f: Y^\om\ra X^\om$ such that $E = f^{-1}(F)$. 
\nl A similar notion exists for the classes of the projective hierarchy: in particular 
A set $F\subseteq X^\om$ is a ${\bf \Si^1_1}$
 (respectively ${\bf \Pi^1_1 }$)-complete set iff for any set $E\subseteq Y^\om$
($Y$ a finite alphabet): 
\nl $E\in {\bf \Si^1_1}$ (respectively $E\in {\bf \Pi^1_1}$) iff there exists a continuous 
function $f$ such that $E = f^{-1}(F)$. 

 A ${\bf \Si^0_\alpha}$
 (respectively ${\bf \Pi^0_\alpha}$)-complete set is a ${\bf \Si^0_\alpha}$
 (respectively ${\bf \Pi^0_\alpha}$)- set which is in some sense a set of the highest 
topological complexity among the ${\bf \Si^0_\alpha}$
 (respectively ${\bf \Pi^0_\alpha}$)- sets. 
\nl  ${\bf \Si^0_n}$
 (respectively ${\bf \Pi^0_n}$)-complete sets, with $n$ an integer $\geq 1$, 
 are thoroughly characterized in \cite{stac}.

 Landweber studied first the topological properties of \orl s.
 He proved that every \orl~ is a boolean combination of $G_\delta$-sets.
and  he also characterized the \orl s in each of the Borel classes 
 ${\bf F, G, F_\sigma, G_\delta }$, and showed that one can decide, for an effectively given
\orl~ $L$, whether $L$ is in  ${\bf F, G, F_\sigma}$, or  ${\bf G_\delta }$.
\nl It turned out that an \orl~ is in the class ${\bf G_\delta }$  iff it 
is accepted by a deterministic B\"uchi automaton. 

 When considering $\om$-CFL, natural questions now arise: are all $\om$-CFL
Borel sets of finite rank, Borel sets, analytic sets....?

 First recall the following previous result, \cite{sta}:

\begin{The}\label{cfana}
Every $\om$-CFL over a finite alphabet $X$ is an analytic subset of $X^\om$.
\end{The}

We showed  the following  

\begin{The}[\cite{finb}]\label{cfnotbor}
\noi
\begin{enumerate}
\ite[(a)] There exist some  $\om$-CFL which are ${\bf \Si_1^1}$-complete sets 
hence non Borel sets.
\ite[(b)] It is undecidable whether an effectively 
given $\om$-CFL is a Borel set. 
\end{enumerate}
\end{The}

\noi Next the $\om$-CFL exhaust the finite ranks of the Borel hierarchy.

\begin{The}[\cite{fin}]
For each non negative integer $n\geq 1$, there exist ${\bf \Si_n^0}$-complete $\om$-CFL $A_n$
and ${\bf \Pi_n^0}$-complete $\om$-CFL $B_n$.
\end{The}

\noi Cohen and Gold proved that one cannot decide whether an $\om$-CFL is in the class 
${\bf F, G }$ or ${\bf G_\delta}$. We have extended in \cite{fin} 
this result to all classes ${\bf \Si_n^0}$ and
${\bf \Pi_n^0}$, for n an integer $\geq 1$, and next to all Borel classes 
in \cite{finb}. (We say that an $\om$-CFL $A$  
is effectively given when a MPDA accepting 
$A$ is given).

  But the 
question was still open whether there exist some  omega
 context free languages which are Borel sets of infinite (but not finite) rank. 
We shall show below that there exist such  
 omega  context free languages.

\section{Operation " exponentiation of sets "}

\noi In order to construct  omega
 context free languages of every finite rank, we used recent results of J. Duparc  
about the Wadge hierarchy. The Wadge hierarchy of Borel sets is a huge refinement of the 
Borel hierarchy. 
Wadge gave first  a description of this hierarchy \cite{wad} and 
Duparc recently got a new proof of Wadge's results and he gave a normal form of Borel sets, 
i.e. an inductive construction of a Borel set of every given degree \cite{dup95} 
 \cite{dup}. In fact we shall need in this paper only some of his results. So we shall recall 
only these results and refer to \cite{dup95} 
 \cite{dup} for more details. 

 Duparc's proof 
relies on set theoretic operations which are the counterpart of arithmetical operations over 
ordinals needed to compute the Wadge degrees.  
 In fact we shall only use in this paper the operation of exponentiation over 
sets of {\it infinite} words. Moreover we shall consider a slight modification 
of Duparc's  operation 
$A \ra A^\sim$ we introduced in \cite{fin} and which we recall now:

\begin{Deff}
Let $ X_A$ be a finite alphabet and $\tla~\notin X_A$. 
\nl Let $X=X_A\cup \{\tla\}$  and $x$ be a finite or  infinite word over the alphabet $X$. 
 \nl Then  $x^\tla$ is inductively defined by:
\nl $\lambda^\tla =\lambda$,
\nl For a finite word $u\in (X_A\cup \{\tla\})^\star$: 
\nl $(u.a)^\tla=u^\tla.a$, if $a\in X_A$,
\nl $(u.\tla)^\tla =u^\tla$  with its  last letter removed if $|u^\tla|>0$, 
\nl $(u.\tla)^\tla$ is undefined if $|u^\tla|=0$,
\nl and for $u$ infinite:
\nl $(u)^\tla = \lim_{n\in\om} (u[n])^\tla$, where, given $\beta_n$ and $v$ in   $X_A^\star$,
\nl $v\sqsubseteq \lim_{n\in\om} \beta_n \lra  \exists n \fa p\geq n\quad  \beta_p[|v|]=v$.
\end{Deff}

\begin{Rem}
For $x \in X^{\leq \om}$, $x^\tla$ denotes the string $x$, once every $^\tla$ occuring in $x$
has been "evaluated" to the back space operation ( the one familiar to your computer!),
proceeding from left to right inside $x$. In other words $x^\tla = x$ from which every
 interval of the form $" a\tla "$ ($a\in X_A$) is removed. We add the convention that 
$(u.\tla)^\tla$ is undefined if $|u^\tla|=0$, i.e. when the last letter $\tla$ can not be used 
as an eraser (because every letter of $X_A$ in $u$ 
has already been erased by some erasers $\tla$ placed  in $u$). 
Remark that the resulting word $x^\tla$ may be 
finite or infinite.
\end{Rem}

\noi For example if  $u=(a\tla)^n$, for $n\geq 1$, 
$u=(a\tla)^\om$ or $u=(a\tla\tla)^\om$ then $(u)^\tla=\lambda$,
\nl if $u=(ab\tla)^\om$ then $(u)^\tla=a^\om$,
\nl if $u=bb(\tla a)^\om$ then $(u)^\tla=b$,  
\nl if $u=\tla(a\tla)^\om$ or $u=a\tla\tla a^\om$ then $(u)^\tla$ is undefined. 

 We can now define the variant  $A \ra A^\approx$ of the operation of 
exponentiation of  sets:

\begin{Deff}
For $A\subseteq X_A^{\om}$ and $\tla~  \notin X_A$, let $X=X_A\cup \{\tla\}$ and
\nl $A^\approx =\{x\in (X_A\cup \{\tla\})^{\om} / x^\tla\in A\}$.
\end{Deff}

\noi  The following result is then another formulation 
of a property of the operation $A \ra A^\sim$ proved in \cite{dup} and which was applied  
in \cite{fin} to study the $\om$-powers of finitary context free languages.

\begin{The}\label{thedup}
Let $n$ be an  integer $\geq 2$ and  $A\subseteq X_A^\om$ be a ${\bf \Pi_n^0}$-complete
 set. Then $A^\approx$ is a 
${\bf \Pi_{n+1}^0}$-complete subset of $(X_A \cup\{\tla\})^\om$.
\end{The}

\noi We  proved that the class $CFL_\om$ is closed under this operation $A \ra A^\approx$.

\begin{The}[\cite{fin}]\label{tildecf}
 Whenever $A\subseteq X_A^\om$ is an $\om$-CFL, then
 $A^\approx \subseteq (X_A\cup\{\tla\})^\om$ is an $\om$-CFL.
\end{The}

\noi {\bf Proof.} An $\om$-word $\sigma\in A^\approx$ may be considered as an  $\om$-word
$\sigma^\tla\in A$ to which we possibly add, before the first letter $\sigma^\tla (1)$ of
  $\sigma^\tla$
(respectively between two consecutive letters $\sigma^\tla (n)$ and $\sigma^\tla (n+1)$
 of $\sigma^\tla$),
a finite word $v_1$ (respectively $v_{n+1}$ )  where:
\nl for all integers $i\geq 1$, $v_{i}$ 
 belongs to the context free (finitary) 
language $L_3$ generated by the context free grammar with the following production rules:
\nl $S\ra aS\tla S$  with $a\in X_A$,
\nl $S\ra \lambda$ ($\lambda$ being the empty word).
\nl this language $L_3$ corresponds to words where every letter of $X_A$ has been removed 
after using the back space operation. 

\begin{Rem} Recall that 
 a one counter automaton is 
a pushdown automaton with a pushdown alphabet in the form 
$\Ga=\{Z_0, z\}$ where $Z_0$ is  the bottom symbol 
 and always remains at the bottom of the pushdown 
store. And a  one counter language is a (finitary) language which is 
accepted by a one counter 
automaton  by final states. 
It is easy to see that in fact $L_3$ is a  \de one-counter language,
 i.e. $L_3$ is accepted by a \de one-counter automaton. 
And for $a\in X_A$,
 the language $L_3.a$ is also accepted by a deterministic one-counter automaton.
\end{Rem}

\noi Then we can see that 
whenever $A\subseteq X_A^\om$, the \ol~ 
$A^\approx \subseteq (X_A\cup\{\tla\})^\om$
is obtained by substituting in $A$ the language $L_3.a$ for each letter $a\in X_A$, 
where $L_3$ is the CFL defined above.
 
  Let now $A$ be an $\om$-CFL given by 
$A= \bigcup_{i=1}^n U_i.V_i^\om$ where $U_i$ and $V_i$
are context free \la s. Then $A^\approx =\bigcup_{i=1}^n (U'_i).V_i^{'\om}$, where $U'_i$
 (respectively $V'_i$) is obtained by substituting the \la~ $L_3.a$ to each letter $a\in X_A$
 in $U_i$ (respectively $V_i$).
\nl The class CFL is closed under substitution, so $U'_i$ and $V'_i$ are CFL. 
Hence the \ol~ $A^\approx$ is an $\om$-CFL 
because 
$\om -KC(CFL)\subseteq CFL_\om$.
 
 We have also given in \cite{fin} an effective  construction of a 
 MPDA  accepting  the \ol~ $A^\approx \subseteq (X_A \cup\{\tla\})^\om$ 
from a MPDA accepting an  \ol~ $A \subseteq X_A^\om$. Recall now the idea of this 
construction. 

  Let $A$ be an $\om$-CFL which is accepted by a Muller pushdown automaton 
 $\mathcal{A}=(K, X_A, \Gamma, \delta, q_0, Z_0, \mathcal{F})$. 
The \ol~ accepted by $\mathcal{A}$ is 
$L(\mathcal{A})= A= \{  \sigma\in X_A^\om$ / there exists a run r
 of $\mathcal{A}$  on $\sigma$ such that $In(r) \in \mathcal{F} \}$.

 We can construct another MPDA  $\mathcal{A}^\approx$ which accepts 
the \ol~ $A^\approx$ over the 
alphabet $X= X_A \cup\{\tla\}$.
\nl Let us describe  informally the behaviour of the machine $\mathcal{A}^\approx$ 
 when it reads 
an $\om$-word $\sigma\in A^\approx$. Recall that this  word  may be considered as an  
$\om$-word
$\sigma^\tla\in A$ to which we possibly add, before the first letter $\sigma^\tla (1)$ of
  $\sigma^\tla$
(respectively between two consecutive letters $\sigma^\tla (n)$ and $\sigma^\tla (n+1)$
 of $\sigma^\tla$),
a finite word $v_1$ (respectively $v_{n+1}$ )  where
 $v_i$ belongs to the context free   language $L_3$. 

 $\mathcal{A}^\approx$ 
 starts the reading as a pushdown automaton accepting the language $L_3$. 
Then $\mathcal{A}^\approx$ 
 begins to read as $\mathcal{A}$, but at any moment of the computation it may guess 
(using the non determinism) that it reads a finite segment $v$ of $L_3$ which will be erased 
(using the eraser $\tla$).  It reads $v$ using an additional stack letter $E$ which 
permits to simulate a one counter automaton at the top of the stack while keeping 
the memory of the stack of $\mathcal{A}$. Then, after the reading of $v$, 
$\mathcal{A}^\approx$ simulates again   the machine $\mathcal{A}$
and so on.

\section{ $\om$-CFL which are Borel of infinite rank}

 A well known example of ${\bf \Pi_2^0}$-complete \orl~  is 
\nl $B_2=\{ \alpha \in\{0, 1\}^\om  / \exists^{\om}  i\quad \alpha (i)=1\}=(0^\star.1)^\om$,
\nl where  $\exists^{\om} i$ means: " there exist  infinitely many $i$ such that $\ldots$". 
\nl $B_2$ is an  omega context free language    because it is an  \orl.

 We can now  get some ${\bf \Pi_{n+2}^0}$-complete set, for an integer $n\geq 1$, from 
the ${\bf \Pi_2^0}$-complete set $B_2$ by 
applying  $n\geq 1 $ times the operation of exponentiation of sets.

  More precisely, we define, for a set $A\subseteq X_A^{\om}$:
\nl $A^{\approx .0}=A$
\nl $A^{\approx .1}=A^\approx$  and
\nl $A^{\approx .(k+1)}=(A^{\approx .k})^\approx$, 
\nl  where we apply $k+1$ times the operation $A\ra A^\approx$ 
with different new letters 
$\tla_1$, $\tla_2$, $\tla_3$, \ldots , $\tla_{k+1}$.  

 We can now infer from 
Theorems \ref{thedup} and \ref{tildecf}  that, for an integer $n\geq 1$,  
$(B_2)^{\approx .n}$ is an omega context free  language which is a 
${\bf \Pi_{n+2}^0}$-complete subset of 
$\{0, 1, \tla_1,\ldots  , \tla_n\}^\om$. Similarly,  if  
$A\subseteq X_A^\om$ is a ${\bf \Pi_2^0}$-complete regular or context free 
\ol~over the alphabet $X_A$, the \ol~ $(A)^{\approx .n}$ is a 
${\bf \Pi_{n+2}^0}$-complete subset of 
$(X_A \cup \{\tla_1,\ldots  , \tla_n\})^\om$. 

 A way to obtain a Borel set of infinite rank, as we shall show below, 
is to define, for two letters 
a, b in $X_A$,  the supremum of the sets $A^{\approx .i}$:

$$\sup_{i\in \mathbb{N}}A^{\approx .i} = \bigcup_{i\in \mathbb{N}} a^{i}.b.A^{\approx .i} $$ 

\noi But this set is defined over an infinite alphabet, and any omega 
context free \ol~  is defined over a finite alphabet. So we have  first to code 
this set over a finite alphabet. We shall first code every set 
$A^{\approx .n}$. The \ol~ $A^{\approx .n}$ is defined over the alphabet 
$X_A\cup\{\tla_1,\ldots , \tla_n\}$ hence we have to
 code every eraser $\tla_j$ by a finite word 
over a fixed finite alphabet. We shall code the eraser $\tla_j$ by the finite word 
$\alpha.B^j.C^j.D^j.E^j.\beta$ over the alphabet $\{\alpha, B, C, D, E, \beta\}$. 
The reason of the coding we choose will be clear later, when we construct a Muller 
pushdown automaton accepting an \ol~ close to the coding of 
$\sup_{i\in \mathbb{N}}A^{\approx .i}$. In fact this MPDA needs to read four times the 
integer $j$ characterizing the eraser $\tla_j$. 

  Remark first that one can define the  morphism 
$$F_n:~~  (X_A\cup \{\tla_1, \ldots , \tla_n\})^\star  \ra  
(X_A \cup \{\alpha, \beta, B, C, D, E\})^\star$$
\noi by  $F(c)=c$ for each $c\in X_A$ and $F(\tla_j)=\alpha.B^j.C^j.D^j.E^j.\beta$ 
for each integer 
$j \in [1, n]$, where $B, C, D, E, \alpha, \beta$ are new letters not in $X_A$.  
This morphism is  naturally extended  to a continuous function 
$$\bar{F_n}:~~  (X_A\cup \{\tla_1, \ldots , \tla_n\})^\om  \ra 
 (X_A \cup \{\alpha, \beta, B, C, D, E\})^\om$$
\noi Then  $\bar{F_n}((X_A\cup \{\tla_1, \ldots , \tla_n\})^\om )$ is the continuous image by 
$\bar{F_n}$  of the compact set $(X_A\cup \{\tla_1, \ldots , \tla_n\})^\om$, hence it is also 
a  compact set, and a closed subset of $(X_A \cup \{\alpha, \beta, B, C, D, E\})^\om$.
We can now state the following lemma. Its proof is easy and left to the reader. 

\begin{Lem}
Let $A\subseteq X_A^\om$ be a ${\bf \Pi_2^0}$-complete subset of $X_A^\om$. Then for each 
integer $n\geq 1$, the  
\ol~  $\bar{F_n}(A^{\approx .n})$  is a ${\bf \Pi_{n+2}^0}$-complete subset of 
$(X_A \cup \{\alpha, \beta, B, C, D, E\})^\om$.
\end{Lem}

We shall prove now that the supremum of the sets $\bar{F_n}(A^{\approx .n})$ is a Borel set 
of infinite rank.

\begin{Lem}
Let  $A\subseteq X_A^\om$ be a ${\bf \Pi_2^0}$-complete subset of $X_A^\om$. 
Then the set 
$$\sup_{n\geq 1}\bar{F_n}(A^{\approx .n})= 
\bigcup_{n\geq 1} a^{n}.b.\bar{F_n}(A^{\approx .n}) $$  
\noi  is a ${\bf \Si_{\om }^0}$-subset of 
$(X_A \cup \{\alpha, \beta, B, C, D, E\})^\om$ which is  not a Borel set of finite rank. 
\end{Lem}

\proo Assume $A\subseteq X_A^\om$ is ${\bf \Pi_2^0}$-complete. Then the preceding lemma 
implies that, for each $n\geq 1$, the \ol~ $\bar{F_n}(A^{\approx .n})$ is a 
${\bf \Pi_{n+2}^0}$-complete subset of 
$(X_A \cup \{\alpha, \beta, B, C, D, E\})^\om$.  Let $a, b$ be two letters in $X_A$ then  
it is easy to show that,  for each $n\geq 1$, 
the set $a^{n}.b.\bar{F_n}(A^{\approx .n})$ is also 
a ${\bf \Pi_{n+2}^0}$-complete  subset of 
$(X_A \cup \{\alpha, \beta, B, C, D, E\})^\om$  thus 
 
$$\sup_{n\geq 1}\bar{F_n}(A^{\approx .n})= 
\bigcup_{n\geq 1} a^{n}.b.\bar{F_n}(A^{\approx .n}) $$  

\noi is in the class ${\bf \Si_{\om }^0}$ by definition of this class. 

 On the other side this set cannot be a Borel set of finite rank. Because if 
$\sup_{n\geq 1}\bar{F_n}(A^{\approx .n})$ was in the class  ${\bf \Pi_j^0}$,  
for an integer $j\geq 1$, 
then the set 
$$a^n.b.\bar{F_n}(A^{\approx .n}) = \sup_{n\geq 1}\bar{F_n}(A^{\approx .n}) \cap 
(a^n.b.(X_A \cup \{\alpha, \beta, B, C, D, E\})^\om)$$ 
\noi would be also in the class ${\bf \Pi_{j}^0}$, because 
$(a^n.b.(X_A \cup \{\alpha, \beta, B, C, D, E\})^\om)$ is a closed hence 
${\bf \Pi_{j}^0}$-set and the class of ${\bf \Pi_{j}^0}$-subsets of  
$(X_A \cup \{\alpha, \beta, B, C, D, E\})^\om$ is closed under finite intersection. 
But this would lead to a contradiction because we have seen that, for 
$n\geq j$, the set $a^n.b.\bar{F_n}(A^{\approx .n})$ is 
${\bf \Pi_{n+2}^0}$-complete, where $n+2 \geq j+2 > j$ hence it is not a 
${\bf \Pi_{j}^0}$-subset of $(X_A \cup \{\alpha, \beta, B, C, D, E\})^\om$. \ep

 We can not show that the \ol~ $\sup_{n\geq 1}\bar{F_n}(A^{\approx .n})$
is an omega context free language. This is connected to 
the fact that the finitary language 
$$\{B^jC^jD^jE^j~/~ j\geq 1\}$$
is not a context free language. But its complement is easily seen to be context free. 
Then, instead of considering  $\sup_{n\geq 1}\bar{F_n}(A^{\approx .n})$, we can add to this 
\ol~ all $\om$-words in the form $a^n.b.u$ where there is in $u$ 
a segment $\alpha.B^j.C^k.D^l.E^m.\beta$, with $j, k, l, m$ integers $\geq 1$, 
which does not code any eraser, or codes an eraser $\tla_j$ for $j>n$. 
Then  we add to $\sup_{n\geq 1}\bar{F_n}(A^{\approx .n})$ another \ol~ which is 
 of  Borel rank $2$ and the resulting \ol~ will be still of infinite 
rank, but we shall show that it is an omega context free language.

 So we  define now formally this   construction in the 
following way. 

 Define first the following context free finitary languages over the alphabet 
\nl $X^\square=(X_A \cup \{\alpha, \beta, B, C, D, E\}):$

$$L^B = \{ a^n.b.u.B^j~ /~ n\geq 1 \mbox{ and } j>n \mbox{ and } u\in (X^\square)^\star\}$$
$$L^C = \{ a^n.b.u.C^j~ /~ n\geq 1 \mbox{ and } j>n \mbox{ and } u\in (X^\square)^\star\}$$
$$L^D = \{ a^n.b.u.D^j~ /~ n\geq 1 \mbox{ and } j>n \mbox{ and } u\in (X^\square)^\star\}$$
$$L^E = \{ a^n.b.u.E^j~ /~ n\geq 1 \mbox{ and } j>n \mbox{ and } u\in (X^\square)^\star\}$$
$$L^{(B,C)} = \{ u.\alpha.B^j.C^k.D^l.E^m.\beta ~ /~ j, k, l, m\geq 1 
\mbox{ and } j\neq k \mbox{ and } u\in (X^\square)^\star\}$$
$$L^{(C,D)} = \{ u.\alpha.B^j.C^k.D^l.E^m.\beta ~ /~ j, k, l, m\geq 1 
\mbox{ and } k\neq l \mbox{ and } u\in (X^\square)^\star\}$$
$$L^{(D,E)} = \{ u.\alpha.B^j.C^k.D^l.E^m.\beta ~ /~ j, k, l, m\geq 1 
\mbox{ and } l\neq m \mbox{ and } u\in (X^\square)^\star\}$$

 Let now 

$$L = L^B \cup L^C  \cup L^D \cup L^E \cup L^{(B,C)} 
\cup L^{(C,D)} \cup L^{(D,E)} $$

\noi It is easy to show that each of the languages 
$L^B, L^C, L^D,  L^E,   L^{(B,C)},  L^{(C,D)},  L^{(D,E)}$ is a context 
free finitary language thus $L$ is also   context free because the class 
CFL is closed under finite union. Then the \ol~ $L.(X^\square)^\om$ is an 
$\om$-CFL which is an open subset of  $(X^\square)^\om$.

 Remark now that any word in $\sup_{n\geq 1}\bar{F_n}(A^{\approx .n})$ belongs to 
the regular \ol~ 
$$R = a^+.b.(X_A \cup (\alpha.B^+.C^+.D^+.E^+.\beta))^\om $$
\noi because every word has an initial segment in the form $a^n.b$ with $n\geq 1$ and 
the letters $\alpha, B, C, D, E, \beta$ are only used to code the erasers 
$\tla_j$ for $j\geq 1$. 

 Consider now the \ol~ 
$$L.(X^\square)^\om  \cap R$$

\noi An $\om$-word $\sigma$ in this language is a word in $R$ such that  
$\sigma$ has an initial word in the form $a^n.b$, with $n\geq 1$, and  
$\sigma$ contains a segment $\alpha.B^j.C^k.d^l.E^m.\beta$ with $j, k, l, m \geq 1$ 
which does not code any eraser $\tla_j$ or codes such an eraser but with $j>n$. 
Thus this \ol~ is disjoint from the set $\sup_{n\geq 1}\bar{F_n}(A^{\approx .n})$. 
Consider now the \ol:

$$A^\bullet = \sup_{n\geq 1}\bar{F_n}(A^{\approx .n}) \cup [ L.(X^\square)^\om  \cap R ]$$

\noi We can now state the next lemma.

\begin{Lem}\label{A-bul-inf}
Let  $A\subseteq X_A^\om$ be a ${\bf \Pi_2^0}$-complete subset of $X_A^\om$. 
Then $A^\bullet$ is a ${\bf \Si_{\om }^0}$-subset of 
$(X_A \cup \{\alpha, \beta, B, C, D, E\})^\om$ which is not a Borel set of finite rank. 
\end{Lem}

\proo Let  $A\subseteq X_A^\om$ be a ${\bf \Pi_2^0}$-complete subset of $X_A^\om$. 
Then we have already seen that $\sup_{n\geq 1}\bar{F_n}(A^{\approx .n})$ is a 
${\bf \Si_{\om }^0}$-subset of 
$(X_A \cup \{\alpha, \beta, B, C, D, E\})^\om$. On the other side 
it is easy to see, from proposition \ref{pi}, 
 that the \orl~ $R$ is a  ${\bf \Pi_2^0}$-set because 
$$R = (R')^\delta$$
\noi where $R'$ is the finitary (regular) language defined by
$$R' = a^+.b.(X_A \cup (\alpha.B^+.C^+.D^+.E^+.\beta))^+$$

\noi Then the \ol~ 
$$L.(X^\square)^\om  \cap R$$
\noi is the intersection of an open set and of a ${\bf \Pi_2^0}$-set. Thus it is also 
a ${\bf \Pi_2^0}$-set because the class ${\bf \Pi_2^0}$ is closed under finite intersection. 
Then the \ol~ 
$$A^\bullet = \sup_{n\geq 1}\bar{F_n}(A^{\approx .n}) \cup [ L.(X^\square)^\om  \cap R ]$$ 
\noi is a ${\bf \Si_{\om }^0}$-subset of 
$(X_A \cup \{\alpha, \beta, B, C, D, E\})^\om$ because the class ${\bf \Si_{\om }^0}$ 
is closed under finite union. 

 We want now to prove that $A^\bullet$ is not a Borel set of finite rank. 
Assume, on the contrary, that $A^\bullet$ is of finite rank $J$, 
where $J$ is an integer $\geq 1$. Then the intersection of $A^\bullet$ and of the complement 
of $L.(X^\square)^\om  \cap R$ would be the intersection of a ${\bf \Pi_{J+1}^0}$-set and of a 
${\bf \Si_2^0}$ hence ${\bf \Pi_3^0}$-set. Hence 

$$\sup_{n\geq 1}\bar{F_n}(A^{\approx .n}) = A^\bullet \cap (L.(X^\square)^\om  \cap R)^-$$ 

\noi would be a ${\bf \Pi_{k}^0}$-set, with $k=max(3, J+1)$. 
But this is not possible because we know from 
the preceding lemma that 
 $\sup_{n\geq 1}\bar{F_n}(A^{\approx .n})$ is a Borel set of infinite rank. \ep 

 We can now state the following 

\begin{The}\label{A-bul-cfl}
Let $A\subseteq X_A^\om$ be an \orl~ over the alphabet $X_A$. Then the \ol~ 
$$A^\bullet = \sup_{n\geq 1}\bar{F_n}(A^{\approx .n}) \cup [ L.(X^\square)^\om  \cap R ]$$ 
\noi is an $\om$-CFL over the alphabet 
$(X_A \cup \{\alpha, \beta, B, C, D, E\})$. 
\end{The}

\proo  We have already seen that $L.(X^\square)^\om $ is an $\om$-CFL, thus 
$$ L.(X^\square)^\om  \cap R $$
\noi is also an $\om$-CFL because the class of omega context free languages is closed 
under intersection with \orl s, \cite{cg}. 

 Suppose the \orl~ $A\subseteq X_A^\om$ is accepted by the \de Muller automaton  
$\mathcal{A}=(K, X_A,\delta, q_0, \mathcal{F}_\mathcal{A})$ where
$\mathcal{A}'=(K, X_A, \delta, q_0)$ is a FSM and 
$\mathcal{F}_\mathcal{A} \subseteq 2^K$ is the collection of 
designated state sets.

 We shall find a MPDA $\mathcal{B}$ accepting an $\om$-CFL $L(\mathcal{B})$ 
such that 
$$\sup_{n\geq 1}\bar{F_n}(A^{\approx .n}) \subseteq L(\mathcal{B}) \subseteq 
A^\bullet = \sup_{n\geq 1}\bar{F_n}(A^{\approx .n}) \cup [ L.(X^\square)^\om  \cap R ]$$ 

\noi Thus we shall have 

$$A^\bullet = L(\mathcal{B}) \cup [ L.(X^\square)^\om  \cap R ]$$ 

\noi And this will imply that $A^\bullet$ is an $\om$-CFL because the class 
$CFL_\om$ is closed under finite union \cite{cg}.

 It is easy to have  $L(\mathcal{B}) \subseteq R$ because if $L(\mathcal{B'})$ is 
an $\om$-CFL which is not included into $R$ one can  replace it by 
 $L(\mathcal{B})=L(\mathcal{B'}) \cap R $ which is then an $\om$-CFL verifying 
$L(\mathcal{B})=L(\mathcal{B'}) \cap R \subseteq R$.

 Recall  now that 
$$L.(X^\square)^\om  \cap R$$
\noi is the set of {\it all} $\om$-words in $R$ having  
 an initial segment  in the form $a^n.b$, with $n\geq 1$, and  
 containing  a segment $\alpha.B^j.C^k.d^l.E^m.\beta$ with $j, k, l, m \geq 1$ 
which does not code any eraser $\tla_j$ or codes such an eraser but with $j>n$.

 Thus,  in order to define the MPDA $\mathcal{B}$, we have  only to 
 consider   the behaviour of $\mathcal{B}$ when reading $\om$-words in the form 
$$a^n.b.u$$
\noi where $n\geq 1$ and $u \in (X_A \cup (\alpha.B^+.C^+.D^+.E^+.\beta))^\om$ is
such that the letters $\alpha, B, C, D, E, \beta$ in $u$ are only used to code the erasers 
$\tla_j$ for $1\leq j\leq n$. 
 ( In order to simplify our notations, we shall sometimes write in the sequel 
$\tla_j=\alpha.B^j.C^j.D^j.E^j.\beta$ and call eraser either $\tla_j$ or its code 
$\alpha.B^j.C^j.D^j.E^j.\beta$, with $j\geq 1$ ).

 And  we have to find  a MPDA $\mathcal{B}$ 
such that $L(\mathcal{B})$ contains  such a word $a^n.b.u$ if and only if  
$u\in \bar{F_n}(A^{\approx .n})$. 

 So we have to look first at $\om$-words in $\bar{F_n}(A^{\approx .n})$.  In such a 
word $\sigma \in \bar{F_n}(A^{\approx .n})$, there are (codes of) erasers $\tla_1, \ldots, 
\tla_n$. The $\om$-word $\sigma$ is in $\bar{F_n}(A^{\approx .n})$ if and only if 
after  the operations of erasing ( with the erasers 
$\tla_1,..., \tla_n$ ) have been achieved in $\sigma$, then the resulting word 
is in $A$.

 Because of the inductive definition of the sets $A^{\approx .n}$, the operations 
of erasing have to be  done in a good order:  in an $\om$-word 
which contains only the erasers $\tla_1,..., \tla_n$, the first operation of erasing uses the 
last eraser $\tla_n$, then the second one uses the eraser $\tla_{n-1}$, and so on \ldots 

 Therefore these operations satisfy the following properties:
\begin{enumerate}
\ite[(a)]  An eraser $\tla_j$ may only erase letters $c\in X_A$ or other 
erasers $\tla_k$ with $k<j$.

\ite[(b)] Assume that in  a word $u \in \bar{F_n}(A^{\approx .n})$, there is a segment 
$c.v.x$ where $c$ is either in $X_A$ or in the set $\{\tla_1, \ldots, \tla_{n-1}\}$, 
and $x$ is (the code of) an eraser $\tla_k$ which erases $c$ when the operations 
of erasing are successively achieved. Now if there is in the segment $v$ (the code of) 
an eraser $\tla_j$ which erases  $e$, where $e\in X_A$ or $e$
 is (the code of) 
another eraser, then  $e$ must belong to $v$ 
(it is between $c$ and $x$ in the word $u$); moreover the operation 
of erasing using the eraser $\tla_j$ has been achieved before that one using the eraser 
$\tla_k$ and this implies that $k \leq j$.  Thus the integer $k$ must verify:
$$k \leq min [ ~~ j ~/~ \mbox{ an eraser } \tla_j \mbox{ has been used into } v]$$

\end{enumerate}

\noi We can now informally describe the behaviour of the MPDA  $\mathcal{B}$ when reading 
a word $a^n.b.u$ such that the letters $\alpha, B, C, D, E, \beta$  
are only used in $u$ to code the erasers 
$\tla_j$ for $1\leq j\leq n$. 

 After the reading of the  initial segment in the form $a^n.b$, the MPDA  $\mathcal{B}$ 
simulates the Muller automaton $\mathcal{A}$ until it guesses, using the non 
determinism, that it begins to read a segment 
$w$ which contains erasers which really erase and some letters of $X_A$ or some other 
erasers which are erased when the operations 
of erasing are achieved in $u$. 

  Then, using the non determinism, when   $\mathcal{B}$ reads 
a letter $c\in X_A$
it may  guess that this letter will be erased and push it in the pushdown store, 
keeping in memory the current state of the Muller automaton $\mathcal{A}$. 

  In a similar manner when   $\mathcal{B}$ reads 
the  code $\tla_j=\alpha.B^j.C^j.D^j.E^j.\beta$ of an eraser, it may guess that 
this eraser will be erased (by another eraser $\tla_k$ with $k>j$) and then it pushes 
in the store the finite word $\gamma.E^j.\varepsilon$, where $\gamma$, $E$, $\varepsilon$ 
are in the pushdown alphabet.

  But  $\mathcal{B}$ 
may also guess that the eraser $\tla_j=\alpha.B^j.C^j.D^j.E^j.\beta$ will really be used
as an eraser.  If it guesses that the code of $\tla_j$ will be used as an eraser, 
$\mathcal{B}$ has to pop from the top of the pushdown store either a letter of $c\in X_A$ or 
the code $\gamma.E^i.\varepsilon$ of another eraser $\tla_i$, with $i<j$, which 
is erased by $\tla_j$. 

 It would be  easy for $\mathcal{B}$ to check whether $i<j$ when reading the initial segment 
$\alpha.B^j$ of $\tla_j$. 

 But as we remarked in item $(b)$ above, the MPDA $\mathcal{B}$ has also to check 
that the integer $j$ is smaller or equal than every integer $p$ such that 
an eraser $\tla_p$ has been used since the letter $c\in X_A$ or the code 
 $\gamma.E^i.\varepsilon$ was pushed 
in the store.  Then, after having pushed in the pushdown store some letter $x\in X_A$ 
or the code $x=\gamma.E^i.\varepsilon$ of an eraser, and before it pops it from the top 
of the store, $\mathcal{B}$ has to keep in the memory of the stack 
the integer  

$$k = min [ p ~/~ \mbox{ some eraser } \tla_p  \mbox{ has been used since  } x 
 \mbox{ was pushed in the stack }]$$

\noi For that purpose $\mathcal{B}$ pushes the finite word $L_2.S^k.L_1$ in the 
pushdown store ($L_1$ is pushed first, then $S^k$ and the letter $L_2$ are 
 pushed in the stack), where $L_1, L_2$ and $S$ are new letters added to the pushdown 
alphabet. 

 So, when $\mathcal{B}$ guesses that $\tla_j=\alpha.B^j.C^j.D^j.E^j.\beta$ will be 
really used as an eraser, there is at the top  of the stack either a letter $c\in X_A$ 
or a code $\gamma.E^i.\varepsilon$ 
of an eraser which will be erased or a code $L_2.S^k.L_1$. The 
 behaviour of $\mathcal{B}$ is then as follows. 

  Assume first  there is  at the top of the stack a code $L_2.S^k.L_1$.  
Then $\mathcal{B}$ firstly checks that $j\leq k$ when reading the 
segment $\alpha.B^j.C$ of the eraser $\alpha.B^j.C^j.D^j.E^j.\beta$.
\nl If $j \leq k$ holds, then $\mathcal{B}$ pops  completely, using 
$\lambda$-transitions, the word  $L_2.S^k.L_1$ from the top of the stack. 
($\mathcal{B}$ has checked it is allowed to use the eraser $\tla_j$). 

 Then there is now in every case 
at the top of the stack either a letter $c\in X_A$ 
or a code $\gamma.E^i.\varepsilon$ 
of an eraser which will be erased. 
The MPDA $\mathcal{B}$ pops this letter $c$ or the code 
$\gamma.E^i.\varepsilon$ ( having checked that $j > i$ after reading the segment 
$\alpha.B^j.C^j$ of the eraser $\alpha.B^j.C^j.D^j.E^j.\beta$ ).
\nl We have  to consider what is now at the top of the stack and distinguish 
three  cases: 
\begin{enumerate}
\ite If there is now at the top of the stack the bottom symbol $Z_0$, 
 then the MPDA $\mathcal{B}$, after having completely read 
the eraser $\tla_j$, may pursue the simulation of the Muller automaton  $\mathcal{A}$ 
or guesses that it begins to read another segment $v$ which will be erased, hence  
the next letter $c\in X_A$ 
or the next code $\alpha.B^m.C^m.D^m.E^m.\beta$ of the word will be erased and then 
$\mathcal{B}$ pushes the letter $c\in X_A$ or the code $\gamma.E^m.\varepsilon$ 
of $\tla_m$ in the pushdown store.

\ite If there is now at the top of the stack either 
 a letter $c'\in X_A$ or a code $\gamma.E^m.\varepsilon$, then  $\mathcal{B}$ pushes 
the code $L_2.S^j.L_1$ in the pushdown store ( $j$ is then the minimum of the set of 
integers $p$ such that an eraser $\tla_p$ has been used since the letter $c'$ or 
the code $\gamma.E^m.\varepsilon$ has been pushed into the stack).  
\ite If there is now at the top of the stack a code $L_2.S^l.L_1$, then the MPDA 
$\mathcal{B}$ has to compare the integers $j$ and $l$ and to replace 
$L_2.S^l.L_1$ by $L_2.S^j.L_1$ if $j<l$. $\mathcal{B}$ achieves this task  while reading 
the segment $D^j.E^j.\beta$ of the eraser $\alpha.B^j.C^j.D^j.E^j.\beta$.
\nl The MPDA $\mathcal{B}$ pops a letter $S$ for each letter $D$ read. It then 
determines whether $j \leq l$. 
\nl If $j \leq l$ then $\mathcal{B}$ pushes $L_2.S^j.L_1$ when reading the 
segment $E^j.\beta$ of the eraser $\tla_j$.
\nl If $j > l$, then when every letter $S$ of the code $L_2.S^l.L_1$ has been popped, there 
are  $(j-l)$ letters $D$ of the eraser $\alpha.B^j.C^j.D^j.E^j.\beta$ which have not yet been 
read by $\mathcal{B}$. When reading these letters the MPDA  $\mathcal{B}$ pushes 
$(j-l)$ letters $U$ in the stack, where $U$ is  a new letter in the pushdown alphabet. Then, 
when reading the segment $E^j$ of the eraser $\alpha.B^j.C^j.D^j.E^j.\beta$,  the MPDA  
 $\mathcal{B}$ firstly pops $U^{j-l}$ (when reading the first $(j-l)$ letters $E$); Afterwards 
 $\mathcal{B}$ pushes again $S^l$ in the stack when reading the rest of the eraser 
$\alpha.B^j.C^j.D^j.E^j.\beta$.  
\end{enumerate}

\noi When the content of the stack is again  just $Z_0$, the initial stack symbol of the 
MPDA $\mathcal{B}$, then $\mathcal{B}$ may pursue the simulation of the Muller automaton $\mathcal{A}$ or 
guesses it begin to read a new segment which will be erased when the operations of erasing 
will be successively achieved.  \ep

 We can now state our main result:

\begin{The}\label{the-inf}
Let  $A\subseteq X_A^\om$ be a ${\bf \Pi_2^0}$-complete \orl~ over the alphabet $X_A$. 
Then $A^\bullet$ is an omega context free language which is a 
 ${\bf \Si_{\om }^0}$-subset of 
$(X_A \cup \{\alpha, \beta, B, C, D, E\})^\om$ but is not a Borel set of finite rank.
\end{The}

\proo It follows directly from  Lemma \ref{A-bul-inf} and Theorem \ref{A-bul-cfl}  \ep 

 In particular if 
$B_2=\{ \alpha \in\{0, 1\}^\om  / \exists^{\om}  i\quad \alpha (i)=1\}=(0^\star.1)^\om$, 
then $(B_2)^\bullet$ is an omega context free language which is a Borel set of infinite rank. 

 Theorem \ref{the-inf} provides infinitely many such $\om$-CFL over any finite alphabet $X$
of cardinal $\geq 2$, because 
there exist infinitely many ${\bf \Pi_2^0}$-complete \orl s over the alphabet $X$, and 
for such an \orl~ $A$ it holds that 

$$A^\bullet \cap X^\om = a^+.b.A$$

\section{Concluding remarks and further work}

We knew that the class of  omega context free languages exhausts the finite ranks of the 
 Borel hierarchy 
and that there exist some  $\om$-CFL which are analytic but non Borel sets. 
We have proven above that there exist some omega context free languages which are 
Borel sets of infinite rank. 

 It is well known that Turing machines, with a B\"uchi or 
Muller acceptance condition, accept \ol s of every Borel rank $< \om_1^{CK}$, where 
$ \om_1^{CK}$ is the first non recursive ordinal \cite{sta}\cite{mos}\cite{sim}. 
Then the following problem  naturally arises: 
describe the  set of infinite Borel ranks of 
omega context free languages, and in particular find  the ordinal 

$$\sup \{\alpha~/~\mbox{ there exist some Borel  } \om-CFL \mbox{ of rank } \alpha \}$$

\noi which is of course $\leq \om_1^{CK}$.  Unfortunately we cannot reach 
some Borel ranks $>\om$ 
by iterating our operation $A\ra A^\bullet$. In fact one cannot even reach some 
${\bf \Si_{\om }^0}$-complete set, as it will be explained in \cite{find} by 
considering the Wadge degrees of Borel sets.

  Recall that the Wadge 
hierarchy of Borel sets is a great refinement of the  Borel hierarchy. 
We proved in \cite{fina} that the length of the Wadge hierarchy of Borel 
$\om$-CFL is an ordinal greater or equal to the Cantor ordinal $\varepsilon_0$, which is the 
first fixed point of the ordinal exponentiation of base $\om$.
Using the above construction of $A^\bullet$, we have improved this result, showing that 
this length is an ordinal greater than or equal to $\varepsilon_\om$, which is the $\om^{th}$ 
fixed point of the ordinal exponentiation of base $\om$, \cite{find}.

\hs {\bf  Acknowledgements.}   Thanks to the anonymous referees for useful comments on 
a previous version of this paper.

\end{document}